\documentclass[twocolumn,showpacs,preprintnumbers,amsmath,amssymb]{revtex4}

\usepackage{graphicx}
\usepackage{dcolumn}
\usepackage{bm}

\begin{document}


\title{From topological insulators to  superconductors and Confinement}

\author{M. Cristina Diamantini}
\email{cristina.diamantini@pg.infn.it}
\affiliation{%
INFN and Dipartimento di Fisica, University of Perugia, via A. Pascoli, I-06100 Perugia, Italy
}%

\author{Pasquale Sodano}
\altaffiliation[Also at ]{On leave of absence from: INFN and Dipartimento di Fisica, University of Perugia, via A. Pascoli, I-06100 Perugia, Italy}
\email{pasquale.sodano@pg.infn.it}
\affiliation{%
Perimeter Institute of Theoretical Physics 31, Caroline St. North, Waterloo, Ontario N2L2Y5, Canada and International Institute of Physics, Federal University of  Rio Grande do Norte,
Av. Odilon Gomes de Lima 1722, Capim Macio, Natal-RN 59078-400, Brazil
}%

\author{Carlo A. Trugenberger}
\email{ca.trugenberger@InfoCodex.com}
\affiliation{%
SwissScientific, chemin Diodati 10, CH-1223 Cologny, Switzerland
}%


\date{\today}

\begin{abstract}

Topological matter in 3D is characterized by the presence of a topological BF term in its long-distance effective action. We show that, in 3D, there is another marginal term that must be added to the action in order to fully determine the physical content of the model. The quantum phase structure is governed by three parameters that drive the condensation of topological defects: the BF coupling, the electric permittivity and the magnetic permeability of the material. For intermediate levels of electric permittivity and magnetic permeability the material is a topological insulator. We predict, however, new states of matter when these parameters cross critical values: a topological superconductor when electric permittivity is increased and magnetic permeability is lowered and a charge confinement phase in the opposite case of low electric permittivity and high magnetic permeability. Synthetic topological matter may be fabricated as 3D arrays of Josephson junctions. 

\end{abstract}
\pacs{11.10.-z,11.15.Wx,73.43.Nq,74.20.Mn}

\maketitle

\section{Introduction}
The last decade has seen a surprising development. Topological field theories \cite{birmi}, originally invented as background-independent theories of quantum gravity, have found a beautiful realization in strongly correlated condensed matter states.

The general idea, pioneered by Wen \cite{Wen}, is to write the conserved matter current describing the fluctuations about such topologically ordered states as the curl of a gauge field and to formulate the low-energy effective field theory for the fluctuations as a gauge theory. In two spatial dimensions, the dominant gauge action at long distances is the famed Chern-Simons term \cite{jackiw, witten}, which describes incompressible quantum Hall fluids and their edge excitations.

Based on exactly the same idea we showed \cite{dst,nsm} how effective field theories with a topological BF term \cite{birmi} provide a unified description of topological matter in any space-time dimensions. The relevance of the BF term for  topological insulators has been recently derived in \cite{topins,moore}. 

The BF term \cite{birmi} is a $P$- and $T$-conserving, higher-dimensional generalization of the Chern-Simons terms. In two spatial dimensions it reduces to a mixed Chern-Simons term, which, for non-compact gauge groups, can be diagonalized into the sum of two standard Chern-Simons terms of opposite chiralities: such theories have often been referred to as doubled Chern-Simons models in the literature \cite{dst,nsm,doubled}. In two spatial dimensions, the topological Chern-Simons term dominates the long-distance effective theory. In three spatial dimensions, instead, there is another marginal term that co-determines the infrared behavior and must be added to the action: this is the standard Maxwell term. 

What distinguishes the different possible phases of this model is the role of the topological defects ensuing from the compactness of the U(1) gauge group. The compactness of this group and its consequences are often overlooked in the literature on this subject. It requires an unavoidable UV cutoff in the formulation of the model  \cite{polbo} and it is this UV cutoff that prevents the diagonalization of compact mixed Chern-Simons terms \cite{nsm}. 

In this paper we show that the $T = 0$ quantum phase structure of topological matter is determined by three parameters: the electric permittivity and the magnetic permeability of the material and the topological coupling constant.  For intermediate levels of electric permittivity and magnetic permeability the material is a topological insulator \cite{topins}. We predict, however, new states of matter when these parameters cross critical values: a topological superconductor when electric permittivity is increased and magnetic permeability is lowered and a dual charge confinement phase in the opposite case of low electric permittivity and high magnetic permeability. 

Our results raise the possibility of turning a topological insulator into a superconductor, if a mechanism can be devised to lower the magnetic permeability and increase the electric permittivity,  while preserving the topological entanglement of the ground state. Equally interesting would be the realization of a charge confinement phase  for high magnetic permeability and low electric permittivity. We also point out that these results can be studied in synthetic BF matter, constituted by fabricated three-dimensional Josephson junction arrays.

Let us write the conserved charge fluctuations above the topological ground state in terms of a Kalb-Ramond \cite{kalb} two-form gauge field $b_{\mu \nu}$  
as  $j_{\mu} = (k/2\pi) \epsilon_{\mu \nu \alpha \beta}\partial_{\nu}b_{\alpha \beta}$. We consider also an effective statistical  gauge field $a_{\mu}$ describing the braiding of charged particles and vortex lines. The corresponding field strength is the pseudo-tensor  $\phi_{\mu \nu} = (k/2\pi)  \epsilon_{\mu \nu \alpha \beta} \partial_{\alpha} a_{\beta}$ describing the current of vortex excitations. Then the BF term is the topological action $ {k \over 2\pi} \int d^4x\ a_{\mu} \epsilon_{\mu \nu \alpha \beta} \partial_{\nu} b_{\alpha \beta}$. Since $a_{\mu}$ and $b_{\mu \nu}$ have canonical dimensions $[mass]$ and $[mass]^2$ (we use units $c = \hbar = 1$), respectively, $k$ is the dimensionless  BF coupling.  In applications to topological matter the BF term represents the effective action for the matter degrees of freedom, obtained by integrating out all microscopic interactions, including electromagnetism.

Contrary to the case of two spatial dimensions, however, in three spatial dimensions, the topological BF term is not the only action term dominating the infrared behavior. Another marginal term can be added to the action: this is a Maxwell-like term constructed from the effective gauge field $a_{\mu}$. We shall thus consider condensed matter systems with the following long-distance Euclidean effective action:
\begin{equation}
S = {ik \over 2 \pi} \int d^4x a_{\mu} \epsilon_{\mu \nu \alpha \beta} \partial_{\nu} b_{\alpha \beta} +  \int d^4x
\left( {1\over 2 e^2 \lambda} {\bf e}^2 +  {1\over 2 e^2 \eta} {\bf b}^2 \right) \ ,
\label{one}
\end{equation}
where ${\bf e}$ and ${\bf b}$ are the effective electric and magnetic fields constructed from the gauge field $a_{\mu}$, $e$ is the electron charge and $\eta$ and $\lambda$ are two dimensionless couplings.  The effects of possible topological term $\propto  \int d^4x {\bf e} \cdot {\bf b}$ will be discussed elsewhere. 

In what follows we will also add to the action an infrared-irrelevant kinetic term for $b_{\mu \nu}$:
\begin{equation}
S \rightarrow S + \int d^4x \left[ {1 \over 2 g^2 \eta } f_0 f_0 +  {1 \over 2 g^2 \lambda } f_i f_i  \right] \ ,
\label{due}
\end{equation}
with $f_\mu \equiv {1 \over 6} \epsilon_{\mu \nu \alpha \beta}f_{\nu \alpha \beta}$,  
$f_{\mu \nu \rho} = \partial_\mu b_{\nu \rho} + \partial_\nu b_{ \rho \mu} +\partial_\rho b_{\mu \nu }$ the Kalb-Ramond field strength, and $g$ a parameter with the dimensions of a mass.
This serves only as an ultraviolet  regulator for Gaussian integrals. It describes the massive modes that live on short length scales $ \ll {1 \over m} = {\pi \over k e g \sqrt{\lambda \eta}}$.

First of all let us analyze the meaning of the parameters $\lambda$ and $\eta$. To this end we introduce the coupling with an external electromagnetic field: 
$i  j_\mu A_\mu=   {i k \over 2 \pi}  A_\mu \epsilon_{\mu \nu \alpha \beta}\partial_\nu b_{\alpha \beta }$,
 in (\ref{one}) 
Gaussian integration over $a_\mu$ and  $b_{\alpha \beta }$ induces the electromagnetic effective action. At long distances ($\gg {1 \over m} $) this is given by:
\begin{equation}
S_{AE} = \int d^4x  {1 \over 2 e^2}\left[ {1\over \lambda} {\bf E}^2 + {1 \over \eta} {\bf B}^2 \right] \ ,
\label{rcee}
\end{equation}
where  ${\bf E}$ is the physical electric field and ${\bf B}$ the magnetic one.
This shows that the model $(\ref{one})$ describes a topological insulator and identifies the parameters of the low-energy effective action as the inverse electric permittivity and magnetic permeability, respectively,
\begin{equation}
\epsilon =  {1  \over \lambda}  \ , \quad  \mu =   \eta \ .
\label{epmu}
\end{equation}

One of the main characteristics of topological insulators is the appearance  of the so-called axion-electrodynamics term \cite{essin,zhang}:
\begin{equation}
S  = \int d^4x {i  \theta  \over 16 \pi^2} F_{\mu \nu}\tilde F^{\mu \nu} = \int d^4x {i  \theta  \over 4  \pi^2} {\bf E} \cdot {\bf B} \ ,
\label{axel}
\end{equation}
in their action when time-reversal symmetry is broken, the parameter $\theta$ being quantized in odd multiples of $\pi$:   $\theta = (2n + 1) \pi \ , n \in  \mathbb{ Z}$  \cite{essin}. $F_{\mu \nu}$ is the field strength of $A_\mu$ and $\tilde F^{\mu \nu}$ its dual.

We will now show how this term arises naturally in our model due to a (T-symmetry breaking) direct coupling of the statistical gauge field to electromagnetic fields. This is represented by the additional term ${i\phi\over 16 \pi} \phi_{\mu \nu} F_{\mu \nu}$ in the action (\ref{one}). Gaussian integration over the matter fields $a_\mu$ and $b_{\mu \nu}$ induces, at long distances ($\gg  {1\over m}$), the axion electromagnetic term with parameter $\theta = { k  \phi \over  2}$. Requiring a T- invariant partition function $\exp ({\rm -S})$ is equivalent to the Dirac quantization condition ${k  \phi  \over   2 \pi} =$ integer. Note that the BF coupling $k$ plays the same role as the charge of the condensate in spontaneous symmetry breaking. As we show below $k$ is indeed an integer.

The action (\ref{one}) has two $U(1)$ gauge symmetries under the transformations:
\begin{equation}
a_{\mu} \to a_{\mu} + \partial_{\mu} \xi\ , 
b_{\mu \nu} \to b_{\mu \nu} + \partial_{\mu}\xi_{\nu} - \partial_{\nu}\xi_{\mu} \ .
\label{two}
\end{equation}
The important point is that these gauge symmetries are compact and the compactness of the $U(1)$ gauge group, as usual, leads to the presence of topological defects. As is beautifully explained in \cite{polbo}, the compactness of  $U(1)$ gauge groups inevitably introduces a length scale, the compactification radius of the gauge fields. There are essentially only two ways to introduce this UV cutoff in a gauge invariant manner: either one embeds the $U(1)$ group in a larger, non-Abelian compact group and one breaks the symmetry down to $U(1)$ on a given scale or one formulates the entire theory on a lattice. We choose the latter.

The lattice we consider is a hypercubic lattice in four Euclidean
dimensions, with lattice spacings $l$.
On this lattice we define the following forward and backward 
derivatives and shift operators: 
\begin{eqnarray}
d_{\mu } f(x) &\equiv {{f(x+l \hat \mu )-f(x)}\over
l}\ ,\  S_{\mu }f(x) \equiv f(x+l \hat \mu )\  , \nonumber \\
\hat d_{\mu } f(x) &\equiv {{f(x)-f(x-l \hat \mu )}\over l} \ ,
\  \hat S_{\mu }f(x) \equiv f(x-l \hat \mu ) \ .
\label{deri}
\end{eqnarray}
Summation by parts interchanges the two derivatives, with a minus sign,
and the two shift operators.
We also introduce the three-index lattice operators \cite{dst}:

\begin{equation}
K_{\mu \nu \alpha} = S_\mu \epsilon_{\mu \rho \nu \alpha} d_\rho \ , \quad \hat K_{\mu \nu \alpha} =  \epsilon_{\mu \nu
 \rho \alpha} \hat d_\rho \hat S_\alpha \  .
 \label{opk}
 \end{equation}
These operators are gauge-invariant in the sense
that:
\begin{eqnarray}
K_{\mu \nu \alpha} d_\alpha &= K_{\mu \nu \alpha}
d_\nu = \hat d_\mu K_{\mu \nu \alpha} = 0 \ , \nonumber \\
\hat K_{\mu \nu \alpha} d_\alpha &= \hat d_\mu \hat K_{\mu \nu
\alpha} = \hat d_\nu \hat K_{\mu \nu \alpha} = 0\ .
\label{pdk}
\end{eqnarray}
Moreover they satisfy the equations:
\begin{eqnarray}
&\hat K_{\mu \nu \alpha} K_{\alpha \lambda \omega} =
K_{\mu \nu \alpha} \hat K_{\alpha \lambda \omega} \equiv \nonumber \\
&\equiv O_{\mu \nu
\lambda \omega } =
 - \left( \delta_{\mu \lambda} \delta_{\nu \omega} - 
\delta_{\mu \omega} \delta_{\nu \lambda}\right) d_\rho \hat d_\rho +\nonumber\\ &+ \left( 
\delta_{\mu \lambda} d_\nu \hat d_\omega - \delta_{\nu \lambda}
d_\mu \hat d_\omega \right) + \left( 
\delta_{\nu \omega} d_\mu \hat d_\lambda - \delta_{\mu \omega}
d_\nu \hat d_\lambda \right) \ ,\nonumber \\
&\hat K_{\mu \omega \alpha} K_{\omega \alpha \nu } =
K_{\mu \omega \alpha} \hat K_{\omega \alpha \nu } \equiv  \nonumber \\
&\equiv 2 D_{\mu \nu} = - 
2 \left(  \delta_{\mu \nu} d_\rho \hat d_\rho - d_\mu \hat d_\nu \right) \ .
\label{ker} 
\end{eqnarray}
The expressions $O_{\mu \nu \lambda \omega }$ and $D_{\mu \nu}$ are 
lattice versions of the Kalb-Ramond and Maxwell kernels, respectively.

On the lattice, the gauge fields $a_\mu$ and $b_{\mu \nu}$ become link and plaquette variables, respectively. The compactness of the gauge groups implies that they are angular variables, invariant under the shifts:
\begin{equation}
a_\mu \rightarrow a_\mu + {2 \pi \over l} n_\mu \ , b_{\mu \nu} \rightarrow b_{\mu \nu} + {2 \pi \over l^2} n_{\mu \nu} \ ;  n_\mu, n_{\mu \nu}  \in  \mathbb{ Z} \ ,
\label{per}
\end{equation}
where $ {2 \pi \over l}$ is the radius of the $U(1)$ group. 
In order to implement this invariance one introduces in the Euclidean partition function integer link and plaquette variables $Q_\mu$ and
$M_{\mu \nu}$ as follows:
\begin{eqnarray}
&Z = \sum_{ \{ Q_{\mu } \} \atop \{ M_{\mu \nu} \} }
\int {\cal D}a_{\mu } \int {\cal D}b_{\mu \nu} \ {\rm exp} (-S) \ ,\nonumber \\
&S = \sum_x { l^4\over 4e^2 \lambda}
 \tilde f_{ij}\tilde f_{ij} + { l^4\over 4e^2 \eta}
 \tilde f_{0i}\tilde f_{0i} +  {l^4 \over 2g^2 \eta} f_0 f_0 +  \nonumber \\
&+ {l^4 \over 2g^2 \lambda} f_i f_i  - i {l^4k \over 2\pi }
a_{\mu }K_{\mu \alpha \beta }b_{\alpha \beta } + \nonumber  \\  
&+ i k \left( l Q_\mu a_\mu \right)  + i k
\left( l M_{\mu \nu} b_{\mu \nu} \right) \ ,
\label{tsbd}
\end{eqnarray}
where $\tilde f_{\mu \nu } \equiv \hat K_{\mu \nu \alpha }
a_{\alpha }$ and  $f_{\mu } \equiv  {1\over 2} K_{\mu \alpha \beta } b_{\alpha \beta }$ .

The integer-valued variables $Q_{\mu }$
and $M_{\mu \nu }$ appearing in (\ref{tsbd}) represent topological excitations whose
role is to make the $BF$ term {\it  periodic}. They satisfy $\hat d_{\mu } Q_{\mu }  = 0$ and $\hat d_{\mu } M_{\mu \nu } = \hat d_{\nu } M_{\mu \nu } =0$. In  Euclidean space they represent closed electric lines and compact magnetic surfaces. The corresponding physical excitations in Minkowski space are the world-lines of quasiparticle charges $Q_0$ and the world-surfaces of magnetic vortex lines $M_{0i}$. 
These topological excitations are  the singularities in the two field strengths due to the compactness of the two gauge symmetries. It can be shown \cite{scs} that they represent chargeons and spinons in 3D.
As we now show, the physics of a compact, $U(1)$ BF term is much richer than that of its non-compact cousin with gauge group $\mathbb{ R}$. 

The phase structure of our model  is determined by the statistical mechanics of a coupled
gas of lattice loops and surfaces. Its partition function can be easily
obtained by a Gaussian integration over the gauge fields $a_{\mu }$ and
$b_{\mu \nu }$ in (\ref{tsbd}):  
\begin{eqnarray}
&Z_{\rm Top} = \sum_{ \{ Q_{\mu } \} \atop \{ M_{\mu \nu} \} } \ {\rm exp}\left( -S_{\rm Top} \right) \ , \nonumber \\
&S_{\rm Top} = \sum_x {e^2 k^2  \lambda \over 2l^2} \ Q_0{1
\over {m^2- \nabla ^2 }} Q_0 +  {e^2 k^2 \eta \over 2l^2} \ Q_i{1
\over {m^2- \nabla ^2 }} Q_i  +Ê\nonumber \\
 &+  g^2  k^2 \lambda  \ M_{0i}
{1 \over {m^2-\nabla ^2}} M_{0i} + g^2  k^2 \eta \ M_{ij}
{1 \over {m^2-\nabla ^2}} M_{ij} \nonumber \\
&+ i {\pi k m^2 \over l}
\ Q_\mu {K_{\mu \alpha \beta }
\over {\nabla ^2 \left( m^2- \nabla^2 \right) }} M_{\alpha \beta}  \ ,
\label{ectop}
\end{eqnarray}
wth the mass $m$ defined by $
m^2 = {e^2g^2 \lambda \eta k^2 \over \pi^2}$ and $\nabla^2 =  (\lambda / \eta)  d_0 \hat d_0  + d_i \hat d_i$ the lattice Laplacian.
Here $\sqrt{\lambda / \eta} = 1/\sqrt{\epsilon \mu}$ is the speed of light in the topological medium.
This partition function can be interpreted as the Euclidean partition
function for a lattice model of interacting particles (whose world-lines are
parametrized by the closed loops $Q_{\mu }$) and closed vortex
strings (whose world-sheets are parametrized by the compact surfaces
$M_{\mu \nu }$). 
By choosing the analogue of the Landau Ginzburg parameter $\alpha=ml \ge O(1)$
the Yukawa interactions in (\ref{ectop}) reduce essentially to contact terms expressing the self-interactions of topological defects. If the  imaginary terms in the action are absent, the model (\ref{ectop}) can be treated by standard statistical mechanics techniques. This is realized when $k$ its an integer, $k = n$, $n \in \mathbb{ Z}$. 

In order to derive the quantum phase structure of the model at $T = 0$ we consider the free energy of static ($Q_i = M_{ij} = 0$) charges and vortex lines. This is given by:
\begin{equation}
F =  \sum_x\left[  {k^2 e^2 \lambda \over \alpha^2 } {Q_0}^2 + {\pi^2 \over  e^2 \eta} {M_{0i }}^2 \right] - \left[ S_Q + S_M \right] \ ,
\label{fren}
\end{equation}
where $S_Q$ and  $S_M$ denote the entropies of closed lines and compact surfaces on the lattice, respectively.These are proportional to the length of world-lines (for quasi-particles) and assumed as proportional to the area of world-surfaces for vortex lines. The exact form of their entropy is not known although these are the well-defined confining strings introduced by Polyakov \cite{pcs}. We can thus derive only qualitative results.

For $k^2 \lambda$ small, the entropy of electric strings dominates their self-energy and the electric part of the free energy is minimized  by configuration in which long electric strings condense. The contrary is true for large $k^2 \lambda$ in which case electric strings are dilute. The same reasoning indicates the condensation of magnetic surfaces for large values of $\eta$, whereas they are dilute for small $\eta$.

All together we find an electric condensation phase for $k^2 \lambda \ll \alpha^2 /e^2$ and  $\eta \ll \pi^2 /e^2$ and a magnetic condensation phase for $k^2 \lambda \gg \alpha^2 /e^2$ and  $\eta \gg \pi^2 /e^2$. In between we have the original topological insulator phase, in which both types of topological excitations are dilute. 

In order to distinguish the various phases we introduce the typical
order parameters of lattice gauge theories \cite{polbo}  namely the {\it Wilson loop} for
an electric charge $q$ and the {\it 't Hooft surface} for a vortex line with flux $\phi $:
\begin{equation}
L_W \equiv {\rm exp}\  i q \left(
\sum_x l j_\mu^{\rm ex} a_\mu \right) \ ,\  \
S_H \equiv {\rm exp}\  i \phi \left( \sum_x
l \phi _{\mu \nu }^{\rm ex} b_{\mu \nu } \right) \ , 
\label{wlts}
\end{equation}
where $j_\mu^{\rm ex}$vanishes  everywhere but on the links of a
closed loop, where it takes the value 1, and where $\phi _{\mu \nu }^{\rm ex}$ vanishes everywhere but on the plaquettes of a  
compact surface, where it takes the value 1.  Since the loops are closed and the surfaces compact
they satisfy
$\hat d_{\mu }j_\mu^{\rm ex}=\hat d_{\mu }\phi _{\mu \nu}^{\rm ex} =0$.

The expectation values of $\langle L_W \rangle $ and $\langle S_H \rangle $ can
be used to characterize the various phases. These expectation values are given by
\begin{eqnarray}
&\langle L_W \rangle = {{Z_{\rm Top} \left( Q_{\mu } +
{q\over k} j_\mu^{\rm ex}, M_{\mu \nu} \right) } \over {Z_{\rm Top} \left(
Q_{\mu }, M_{\mu \nu} \right) }} \ ,\nonumber \\
&\langle S_H \rangle = {{Z_{\rm Top } \left( Q_{\mu }, M_{\mu \nu}
+ { \phi \over k }
\phi _{\mu \nu}^{\rm ex} \right) } \over {Z_{\rm Top} \left( Q_{\mu },
M_{\mu \nu} \right) }} \ ,
\label{exva}
\end{eqnarray}
where the notation is self-explanatory.  The dominant contributions are given by \cite{dst}
\begin{eqnarray}&\langle L_W \rangle _{\rm m.\ c.} =
\sum_x {e^2q^2  \lambda \over 2l^2} \ j_0^{\rm ex}{1
\over {- \nabla ^2 }} j_0^{\rm ex} +  {e^2q^2 \eta \over 2l^2} \ j_i^{\rm ex}{1
\over {- \nabla ^2 }} j_i^{\rm ex}  Ê\nonumber \\
&\langle S_H \rangle _{\rm e.\ c.} = \sum_x g^2  \phi^2 \lambda \ \phi_{0i}^{\rm ex}
{1 \over {-\nabla ^2}} \phi_{0i}^{\rm ex} + g^2 \phi^2 \eta \ \phi_{ij}^{\rm ex}
{1 \over {-\nabla ^2}} \phi_{ij}^{\rm ex} \ .
\label{tdomcom}
\end{eqnarray}
As is evident from these expressions, the effect of the condensation of topological defects is to transform the original short-range Yukawa interactions of chargeons and spinons into long-range Coulomb interactions. The consequences of this are best appreciated by expressing the 't Hooft surface and the Wilson loop in terms of external gauge potentials $A_{\mu}$ and $B_{\mu \nu}$, respectively, 
$\phi_{\mu \nu}^{\rm ex} \propto \hat K_{\mu \nu \alpha} A_{\alpha}$ and $j_{\mu}^{\rm ex}  \propto \hat K_{\mu \alpha \beta} B_{\alpha \beta}$. Then the induced charge and vortex currents can be computed as:
\begin{eqnarray}
j_i^{\rm ind} &&\propto {\delta \over \delta A_i} \langle S_H (A) \rangle _{\rm e.\ c.}  \propto A_i \ ,
\nonumber \\
\phi_{ij}^{\rm ind} &&\propto {\delta \over \delta B_{ij}} \langle L_W (B) \rangle _{\rm m.\ c.} \propto B_{ij} \ .
\label{london}
\end{eqnarray}
These are London equations for the induced charge and spin currents, which express perfect charge conductivity and a photon mass in the electric condensation phase and perfect spinon conductivity, electric screening  and a photon mass \cite{que} in the magnetic condensation phase. We thus conclude that the electric condensation phase is a topological superconductor, whereas the magnetic condensation phase is a charge confinement phase (spinon or dual superconductor).  As  explained above these topological superconductors are characterized by a flux quantization condition $\phi = 2 \pi /k $. The charge confinement phase is the exact dual of the superconducting phase  \cite{hooft}. The photon becomes massive due to the St\"uckelberg mechanism, electric fields are expelled and restricted to thin flux tubes between particle-antiparticles pairs and no charged excitations can be observed.  Note that the original $2 \pi$ periodicity in $\theta$ of topological insulators is lost in the confinement phase. All together this amounts to  the following $T = 0$ quantum phase structure
\begin{eqnarray}
&\epsilon \gg { k^2 e^2 \over \alpha^2}  \ ,  \mu \ll  {\pi^2  \over  e^2}  &\rightarrow {\rm top. \ superconductor} \ ; \nonumber  \\ 
&{\rm intermediate\ regime} &\rightarrow   {\rm top. \ insulator}\ ; \nonumber \\
&\epsilon \ll { k^2 e^2 \over \alpha^2}  \ ,  \mu \gg   {\pi^2  \over  e^2} &\rightarrow {\rm top. \ confinement} \ .
\label{fasi}
\end{eqnarray}

As we have shown in \cite{dst}, topological matter can be simulated in synthetic materials constructed as arrays of Josephson junctions. In what we called the self-dual approximation, such arrays are governed exactly by the model (\ref{tsbd}) with $k = 2$ and:
\begin{equation}
\epsilon =  {e^2\over 4E_C l } \ , \mu \rightarrow \infty \ , \ , g^2 = {\pi^2 e^2 \over 8}{E_J \over  E_C l^2} \ ,
\label{crjo}
\end{equation}
where $E_J$ and $E_C$ are the Josephson coupling and the charging energy of the array, respectively. In this case the phases correspond to the superconducting, metallic and insulating phases of the array. The phase transitions have to be analyzed by varying $E_J$ and $E_C$ at fixed $m l = \sqrt{8 E_C E_J l^2} \ge O(1)$ and are due to a competition between the superconducting junction favoring a global phase lock-in and the charging energy favoring the localization of charges on the islands.

 Note also that a phase structure analogous to (\ref{fasi}) has been found in \cite{ype} in the compact BF formulation of 2D doped Mott insulators.

\end{document}